\documentclass[conference]{IEEEtran}
\IEEEoverridecommandlockouts
\usepackage{cite}
\usepackage{amsmath,amssymb,amsfonts,amsthm}
\usepackage{algorithmic}
\usepackage{graphicx}
\usepackage{textcomp}
\usepackage{xcolor}
\usepackage{cleveref}
\usepackage{relsize}
\usepackage{url}
\usepackage{array}
\usepackage{hhline}

\usepackage{todonotes}
\newcommand{\para}[1]{\vspace{1.5mm}\noindent\textbf{#1}}

\newcolumntype{?}{!{\vrule width 1pt}}

\bstctlcite{IEEEexample:BSTcontrol}

\usepackage{multirow}
\usepackage{tabularx}
\usepackage{tikz}
\usepackage{graphicx}
\usepackage{longtable}
\usepackage{pifont}
\usepackage{xspace}
\usepackage{booktabs}
\usepackage{pgfplotstable}
\usepackage[table]{xcolor}
\usepackage[dvipsnames]{xcolor}
\usepackage[most]{tcolorbox}

\colorlet{mylightgray}{gray!10}
\tcbset{ 
        boxsep=4pt, left=0pt,right=0pt,top=0pt,bottom=0pt,
        colframe=white,colback=mylightgray,  
        highlight math style={enhanced},
        width=\columnwidth
        }

\newtcolorbox{italicnote}{
  enhanced,          
  breakable,         
  colback=mylightgray, 
  colframe=mylightgray, 
  fontupper=\itshape,  
  arc=2mm,           
  boxrule=1pt,       
  boxsep=1mm,        
  left=1mm,          
  right=1mm          
}

\newcommand{\cmark}{\textcolor{green!60!black}{\ding{51}}}  
\newcommand{\xmark}{\textcolor{red!80!black}{\ding{55}}}    
\newcommand{\framework}{Gaia\xspace}
\newcommand{\mybox}[1]
{
\begin{italicnote}
#1
\end{italicnote}
\vspace{1.5mm}
}

\theoremstyle{definition}
\newtheorem{definition}{Definition}

\def\BibTeX{{\rm B\kern-.05em{\sc i\kern-.025em b}\kern-.08em
    T\kern-.1667em\lower.7ex\hbox{E}\kern-.125emX}}
\begin{document}

\title{
The Missing Dimensions in Geo-Distributed Database Evaluation
}

\author{
    \IEEEauthorblockN{Oto Mraz\IEEEauthorrefmark{1}, 
                      Kyriakos Psarakis\IEEEauthorrefmark{2}, 
                      George Christodoulou\IEEEauthorrefmark{1}, 
                      Paris Carbone\IEEEauthorrefmark{3}, and 
                      Asterios Katsifodimos\IEEEauthorrefmark{1}}

    \IEEEauthorblockA{
        \IEEEauthorrefmark{1}\textit{Delft University of Technology,} 
        \IEEEauthorrefmark{2}\textit{Ververica GmbH,} 
        \IEEEauthorrefmark{3}\textit{KTH Royal Institute of Technology}
    }
    
    \IEEEauthorblockA{
        \{o.m.mraz, g.c.christodoulou, a.katsifodimos\}@tudelft.nl, 
        kyriakos.psarakis@ververica.com, 
        parisc@kth.se
    }
}

\maketitle

\begin{abstract}
Geo-distributed OLTP databases are widely deployed across cloud regions, yet current evaluation practices do not cover the challenges of this aspect. Existing benchmarks assume stable network conditions; they lack explicit settings for data and client locality, and they largely ignore data transfer costs across regions. In addition, most evaluations rely on a limited set of geo-distribution patterns. In this paper, we propose \framework, a comprehensive evaluation framework that addresses these gaps. We use \framework to perform a comprehensive evaluation of existing geo-distributed OLTP systems. We deploy them across multiple cloud regions, using different geo-distribution patterns and variable cross-region network conditions. Among other interesting findings, our framework reveals that: $i)$~most systems are sensitive to network instabilities, $ii)$~network costs dominate cloud deployment expenses $iii)$~multi-region fault-tolerance mechanisms incur measurable critical-path overhead that is often overlooked in prior evaluations. We argue that for the design of future geo-distributed databases, we must rethink the trade-offs between performance, fault-tolerance, and cost.
\end{abstract}


\section{Introduction}\label{sec:intro}

Geo-distributed transactional databases are fundamental to modern applications across banking, e-commerce, and financial trading. By partitioning and replicating data across different geographical regions, these systems increase availability, tolerate regional failures, and provide low-latency access to users. However, ensuring strong consistency and atomicity across regions requires complex coordination protocols that may entail multiple communication rounds. Due to wide-area network (WAN) latencies and cross-region communication, transactions may incur significant overhead compared to single-region deployments. Such coordination and concurrency control protocols have been the subject of research in recent years~\cite{nguyen2023detock,harding2017evaluation,ren2019slog,thomson2012calvin,hildred2023caerus,taft2020cockroachdb,faunadb,zhou2025oltp,nguyen2025sunstorm}. 

To manage these trade-offs, many systems adopt deterministic execution models~\cite{thomson2012calvin,ren2019slog}, in which transaction order is determined before execution. Determinism is particularly attractive in geo-replicated settings, where timestamp-based approaches suffer from clock skew and message delays~\cite{BernsteinG81}. Without determinism, conflicting transactions may be aborted or reordered inconsistently, leading to replica divergence. By processing transactions in a predefined global order, deterministic databases enforce convergence and reduce coordination.

\renewcommand{\xmark}{\color{BrickRed!80}{\tiny\ding{108}}} 
\renewcommand{\cmark}{\color{ForestGreen}{\footnotesize\ding{51}}}
\begin{table}[t]
\centering
\rowcolors{5}{White}{gray!30}
\caption{A comparison of the scenarios and metrics used in the evaluations of geo-distributed databases.}
\vspace{-3mm}
\label{tab:evaluation_comparison}
\resizebox{\columnwidth}{!}{%
\begin{tabular}{r?c c c c c c c c}
\specialrule{.1em}{.05em}{.05em}
\multicolumn{1}{c}{} & \multicolumn{6}{c}{\textbf{Evaluation Scenario}} & \multicolumn{2}{c}{\textbf{Metric}} \\ 
\cmidrule(lr){2-7} \cmidrule(lr){8-9} 
\multicolumn{1}{c}{\textbf{Publication}} & \multicolumn{1}{c}{Scalability} & \multicolumn{1}{c}{\begin{tabular}[c]{@{}c@{}}Access\\patterns\end{tabular}} &  \multicolumn{1}{c}{\begin{tabular}[c]{@{}c@{}}Access\\skew\end{tabular}} & \multicolumn{1}{c}{\begin{tabular}[c]{@{}c@{}}Network\\conditions\end{tabular}} & \multicolumn{1}{c}{\begin{tabular}[c]{@{}c@{}}Fault\\tolerance\end{tabular}} & \multicolumn{1}{c}{\begin{tabular}[c]{@{}c@{}}Resource\\allocation\end{tabular}} & \multicolumn{1}{c}{\begin{tabular}[c]{@{}c@{}}Data\\transfers\end{tabular}} & \multicolumn{1}{c}{\begin{tabular}[c]{@{}c@{}}Monetary\\costs\end{tabular}}\\ 
\specialrule{.1em}{.05em}{.05em}
    Calvin~\cite{thomson2012calvin}      & \cmark & \xmark & \cmark & \xmark & \xmark & \xmark & \xmark & \xmark \\ \hline
    SLOG~\cite{ren2019slog}              & \cmark & \cmark & \cmark & \xmark & \xmark & \xmark & \xmark & \xmark \\ \hline
    Detock~\cite{nguyen2023detock}       & \cmark & \cmark & \cmark & \cmark & \xmark & \xmark & \xmark & \xmark \\ \hline
    Caerus~\cite{hildred2023caerus}      & \cmark & \cmark & \xmark & \xmark & \xmark & \xmark & \xmark & \xmark \\ \hline
    Aria~\cite{lu13aria}                 & \cmark & \xmark & \cmark & \xmark & \xmark & \xmark & \xmark & \xmark \\ \hline
    CRDB~\cite{taft2020cockroachdb}      & \cmark & \xmark & \xmark & \xmark & \cmark & \xmark & \xmark & \xmark \\ \hline
    Spanner~\cite{corbett2013spanner}    & \cmark & \xmark & \xmark & \xmark & \xmark & \xmark & \xmark & \xmark \\ \hline
    Aurora~\cite{verbitski2017amazon}    & \cmark & \xmark & \xmark & \xmark & \xmark & \cmark & \xmark & \xmark \\ \hline
    TAPIR~\cite{zhang2018building}       & \cmark & \xmark & \cmark & \xmark & \cmark & \xmark & \xmark & \xmark \\ \hline
    Janus~\cite{mu2016consolidating}     & \cmark & \xmark & \cmark & \xmark & \xmark & \xmark & \xmark & \xmark \\ \hline
    Mako~\cite{shen2025mako}             & \cmark & \cmark & \xmark & \xmark & \cmark & \xmark & \xmark & \xmark \\ \hline
    MaaT~\cite{mahmoud2014maat}          & \cmark & \cmark & \xmark & \xmark & \cmark & \xmark & \xmark & \xmark \\ \hline
    EC~\cite{gupta2018easycommit}        & \cmark & \cmark & \cmark & \xmark & \xmark & \xmark & \xmark & \xmark \\ \hline
    Q-Store~\cite{qadah2020q}            & \cmark & \cmark & \cmark & \xmark & \xmark & \xmark & \xmark & \xmark \\ \hline
    Deneva~\cite{harding2017evaluation}  & \cmark & \xmark & \cmark & \cmark & \xmark & \xmark & \xmark & \xmark \\ \hline
CRDB-MR~\cite{vanbenschoten2022enabling} & \cmark & \cmark & \xmark & \xmark & \xmark & \xmark & \xmark & \xmark \\ \hline
DB sen.~\cite{wang2022study}             & \cmark & \xmark & \cmark & \cmark & \xmark & \xmark & \xmark & \xmark \\ 
\specialrule{.1em}{.05em}{.1em} 
\rowcolor{green!10!white}
{\textbf{\framework}} & \cmark & \cmark & \cmark & \cmark & \cmark & \cmark & \cmark & \cmark \\
\specialrule{.1em}{.05em}{.05em}
\end{tabular}
}
\end{table}

Despite the growing adoption of geo-distributed systems in industry (e.g., Spanner \cite{corbett2013spanner}, Aurora \cite{verbitski2017amazon}, CockroachDB \cite{taft2020cockroachdb}), current benchmarking practices do not capture the intricacies of their deployments. Standard OLTP benchmarks such as TPC-C~\cite{tppc2010tpcc}, YCSB~\cite{cooper2010benchmarking}, and SmallBank~\cite{alomari2008cost} were originally designed for traditional single-region deployments more than a decade ago. As we show in this paper, their schemata and data distributions offer limited support for modeling important geo-distributed transactional patterns. However, more complex geo-distribution patterns would better reflect the variations of real-world workloads. Moreover, most evaluations~\cite{thomson2012calvin,ren2019slog,nguyen2023detock,lu13aria,zhang2018building,mu2016consolidating,harding2017evaluation,wang2022study} focus on scalability and skewed data distribution, but they rarely model failures, unstable networks, or heterogeneous cloud infrastructures (\Cref{tab:evaluation_comparison}). Although WAN communication can be costly in cloud deployments, data transfer volumes and pricing implications are often overlooked in the literature. Given the increasing adoption of geo-distributed databases nowadays, novel benchmarking methods, evaluation scenarios, and metrics have become especially important and timely.

\begin{figure}[t]
\centering
\includegraphics[width=\columnwidth]{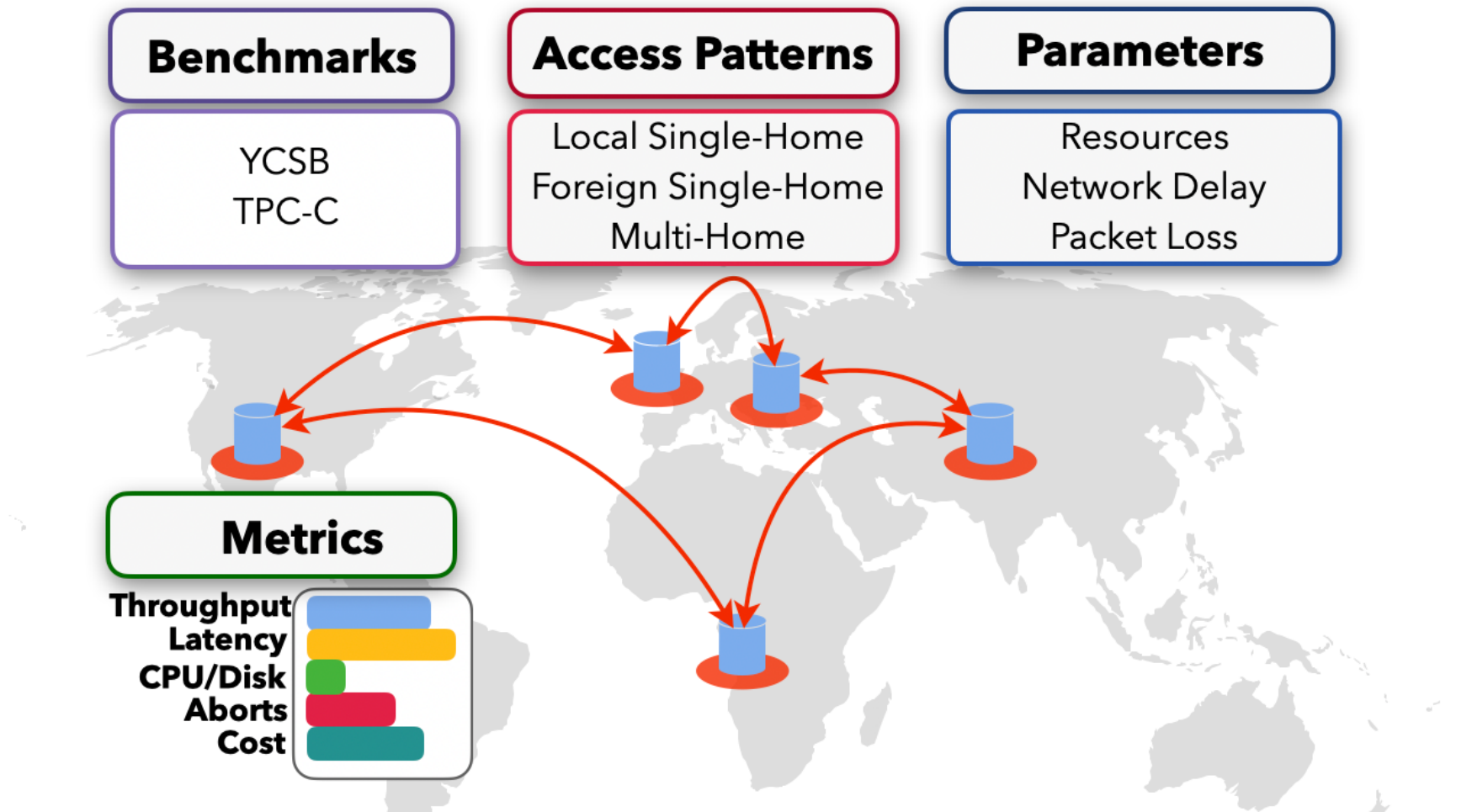}
\caption{\framework framework design. \framework generates transactions based on user specifications and returns performance and resource utilization metrics.}
\label{fig:scenarios:system_architecture}
\end{figure}

To address these issues, we develop \framework, a benchmarking framework tailored to geo-distributed OLTP databases. We use \framework to evaluate existing state-of-the-art geo-distributed databases across scenarios that were not tested in their original publications. \framework integrates a range of OLTP systems and evaluates them under different scenarios derived from existing benchmarks such as TPC-C~\cite{tppc2010tpcc} and YCSB~\cite{cooper2010benchmarking} (\Cref{fig:scenarios:system_architecture}). These scenarios model realistic conditions, including transaction locality, submission-placement asymmetry, hardware instability, and varying hardware configurations. In addition, \framework introduces new metrics, including cross-region data transfers and monetary costs, to provide a complete picture of system behavior under varying deployment conditions. \framework is designed to be extensible, allowing new geo-distributed protocols to be integrated with minimal effort.

Using \framework, we evaluate five geo-distributed OLTP systems deployed across multiple cloud regions. We analyze system performance across eight scenarios and examine how design decisions affect performance and costs in WAN deployments. Note that our goal in this work is not to compare systems but to identify key limitations in current evaluation approaches and propose new metrics and evaluation scenarios. 

\noindent In short, this paper makes the following contributions:

\begin{itemize}
    \item The first geo-distributed database evaluation framework that considers data transfers, and cloud deployment costs alongside the traditional latency and throughput metrics.
    \item A taxonomy of scenarios for geo-distributed transactions.
    \item A comprehensive performance evaluation using our framework of five database systems in different deployment scenarios, yielding new insights on geo-distributed database designs and their tradeoffs.
    \item Identification of high-latency cross-regional data transfers as the primary cost for geo-distributed systems.
    \item Quantification of how network instability affects the performance of existing systems.
    \item Demonstration of the overhead that multi-region fault-tolerance mechanisms impose on transaction performance in geo-distributed systems.
\end{itemize}


The rest of the paper is organized as follows. We first provide an access pattern taxonomy (\ref{sec:txn_types}), summarize geo-distribution modeling in existing benchmarks (\ref{sec:geodistribution}) and system evaluations (\ref{sec:background}). We derive a more holistic set of evaluation scenarios (\ref{sec:scenarios_metrics:eval_scenarios}) and metrics (\ref{sec:metrics}). In (\ref{sec:eval}) we discuss the insights our framework reveals about system capabilities and outline open challenges (\ref{sec:discussion}). (\ref{sec:related_work}) provides a related work overview.

\framework is available at \url{https://github.com/delftdata/gaia}. We present additional metrics in our extended report~\cite{extended_gaia_report}.

\section{Transactional Access Patterns}\label{sec:txn_types}

In addition to multiple works omitting specific benchmarking scenarios, we also observe a lack of critical workload characteristics in the two most popular benchmarks, YCSB~\cite{cooper2010benchmarking} and TPC-C~\cite{tppc2010tpcc}. In this section, we define transaction types and analyze their prevalence in the state-of-the-art evaluation benchmarks.

\begin{figure}[t]
\centering
\includegraphics[width=\columnwidth]{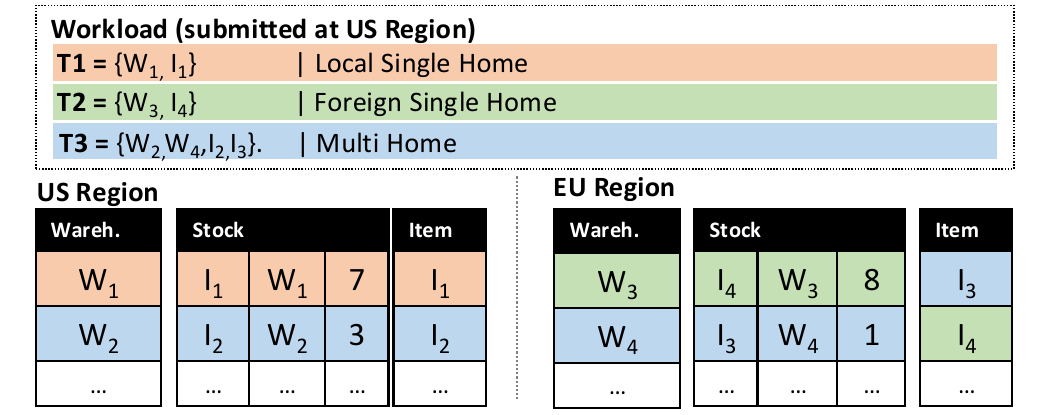}
\caption{Three transaction types distinguished by \framework (replicas omitted).}
\label{fig:transaction_types}
\end{figure}

Similar to~\cite{ren2019slog,nguyen2023detock,hildred2023caerus}, \framework, classifies transactions based on two properties that determine their access patterns: $i)$ the number of partitions accessed and $ii)$ the number of regions involved. We distinguish between: \emph{Single-Partition (SP) transactions}, which access data within a single partition, and \emph{Multi-Partition (MP) transactions}, which access data in two or more partitions. However, unlike prior work which classifies transactions as single-home or multi-home, \framework refines this categorization into three distinct classes based on the location of the \textit{client initiating} the transaction. The different transaction types are summarized in \Cref{tab:transaction_types}. \Cref{fig:transaction_types} illustrates the differences between these three classes on a simplified TPC-C schema. Let $R(t)$ denote the set of regions involved in a transaction $t$, and let $O(t)$ be the origin region of $t$, i.e., the region where the transaction was submitted. We distinguish the following three types of transactions:

\begin{definition}[Local Single Home (LSH)]
A transaction is considered LSH if the primary replica of all data objects it accesses resides within a single region, and this region matches the transaction's region of origin. LSH transactions do not require cross-region round-trips as the data is located in the same region as the query. Formally: 
    $
    \quad |R(t)| = 1 \quad \text{and} \quad O(t) = R(t).
    $
\end{definition}

\begin{definition}[Foreign Single Home (FSH)]
An FSH transaction also accesses the primary replica of all data objects withing a single region, but that region differs from the region where the transaction was submitted. FSH transactions require the forwarding of the transaction to the respective home region of the data and sending back the result to the region of origin (i.e., at least one round trip is required). Formally:
    $
    \quad |R(t)| = 1 \quad \text{and} \quad O(t) \neq R(t).
    $
\end{definition}   

There are two key reasons for distinguishing FSH and LSH: $i)$~they exhibit different communication patterns during transaction processing. An FSH transaction is submitted to a region that differs from the region where the data is actually mutated. FSH transactions may thus incur higher latency than LSH, since an additional round-trip is required for data collection and, $ii)$~network costs are involved when exchanging data across the two regions for FSH.

\begin{definition}[Multi Home (MH)]
A transaction is MH  if it accesses the primary copies of data across two or more regions and requires coordination across regional database instances. Formally:
    $
    \quad |R(t)| \geq 2.
    $
MH transactions require at least one round trip to send the transaction and gather the results. In addition, they may require additional rounds to coordinate with other regions involved in the transaction to resolve potential concurrency conflicts.\end{definition} 

\para{Partitions vs. Homes.} Note that a given benchmark assigns a set of partitions to each region. By default, the home of a given \textit{primary} partition is assigned randomly and uniformly. A partition may be replicated in other regions, but this does not affect the definition of LSH, FSH, and MH transactions: coordination occurs across primary (also called master) replicas, not secondary replicas.

\begin{table}[t]
\centering
\rowcolors{2}{White}{gray!30}
\caption{Transactional access patterns distinguished by \framework.}
\label{tab:transaction_types}
\vspace{-3mm}
\resizebox{\columnwidth}{!}{%
\begin{tabular}{ll}
\bottomrule
\textbf{Name}                                                     & \textbf{Transactional Access Pattern}                                                                                                                             \\\specialrule{.1em}{.05em}{.05em}
Single-Partition (SP)                                                & Within a single partition                                                                                                  \\ \hline
Multi-Partition (MP)                                                 & Across two or more partitions                                                                                                  \\ \hline
\begin{tabular}[c]{@{}l@{}}Local Single-Home (LSH)\end{tabular}   & \begin{tabular}[c]{@{}l@{}}Within a single region (the one where \\ the transaction originates)\end{tabular}              \\ \hline
\begin{tabular}[c]{@{}l@{}}Foreign Single-Home (FSH)\end{tabular} & \begin{tabular}[c]{@{}l@{}}Within a single region (different from \\ the transaction's origin)\end{tabular} \\ \hline
Multi-Home (MH)                                                      & Across two or more regions                                                                                                      \\ \toprule
\end{tabular}%
}
\end{table}

\para{Discussion.} In real-world systems, challenges arise from the variations of transactions' data access patterns. For instance, FSH transactions can occur when an e-commerce customer orders goods stored in a warehouse abroad. Systems like Detock~\cite{nguyen2023detock} address this behavior by implementing \textit{home-movement} functionality. If the system detects that most transactions touching a partition originate from a remote region, it relocates the home of that partition to the originating region. Thus, systems transform FSH transactions into LSH transactions. However, if a particular partition is frequently accessed by FSH transactions originating from multiple regions, relocating its home proves ineffective. For various reasons, including security, network disruptions, and costs, changing the location of data storage can incur additional overhead. 

However, even with reduced cross-region traffic, ensuring atomicity and consistency remains challenging for MP read-write transactions: the system must either commit or abort. Within a single region, coordination is relatively inexpensive, often below 1ms. In a geo-distributed setting, however, inter-region latencies can be up to three orders of magnitude higher, making coordination a performance bottleneck.


\section{Geo-distribution in Existing Benchmarks}\label{sec:geodistribution}

In this section, we discuss the suitability of YCSB~\cite{cooper2010benchmarking} and TPC-C~\cite{tppc2010tpcc} for evaluating different transactional access patterns in geo-distributed settings. At a high level, the suitability of a benchmark for evaluating geo-distributed transactions depends on the shape of their schemas, the distribution of values in their tables, and the locations of the records that transactions touch during execution.

\para{YCSB.} Due to its simplicity, the Yahoo! Cloud Serving Benchmark (YCSB)~\cite{cooper2010benchmarking}, can be adapted for MP and MH transactions. It consists of a single table with a primary key and a varying number of columns, with each column containing 100 random bytes by default. The main benefit of YCSB lies in its flexibility. Additional parameters, such as skew, MH, and MP percentages, are already present in prior work~\cite {ren2019slog,nguyen2023detock}, making it applicable to testing database systems under varying conditions. In \framework, we have included tuning knobs to set the percentages for LSH, FSH, and MH transactions. 

\para{TPC-C and Geo-distribution.} The TPC-C benchmark~\cite{tppc2010tpcc} simulates a realistic OLTP workload that resembles an online ordering and delivery system. The benchmark includes five transactions (\texttt{NewOrder}, \texttt{Payment}, \texttt{OrderStatus}, \texttt{Delivery}, \texttt{StockLevel}) over nine tables. The most interesting transactions are \texttt{NewOrder} and \texttt{Payment} because they can access data from remote regions, thereby inducing cross-regional coordination. Unlike YCSB, the default TPC-C benchmark specifications predefine a fixed transaction type composition. 

The conventional way data is spread across regions to create MH transactions is to place warehouses, and their contained products in separate regions~\cite{nguyen2023detock,harding2017evaluation,ren2019slog,thomson2012calvin,cui2025bonspiel,taft2020cockroachdb,abebe2020dynamast,hildred2023caerus}. Typically, a \texttt{NewOrder} transaction involves an order of 10 items. Each item has a 99\% probability of being sourced from a warehouse in the same region as the client and a 1\% chance to be delivered from a remote warehouse, making the transaction cross-regional. Similarly, each \textit{Payment} transaction defaults to a 1\% chance of involving a remote warehouse. Hence, production workloads with different access patterns, e.g., more MH transactions, cannot be easily modeled by TPC-C. To verify this hypothesis, we analyze the default TPC-C specification in practice and assess how its design choices influence the transactional access patterns. We generate two million TPC-C transactions and report their composition in Table~\ref{tab:tpcc_txn_percentages}.

The breakdown reveals a striking imbalance: over 95\% of transactions are LSH, around 4\% are MH, and $\leq0.5$\% are FSH. Interestingly, all observed FSH transactions correspond to \texttt{Payment} operations that update a single warehouse issued from a remote warehouse. In theory, \texttt{NewOrder} transactions could also be FSH if all ten ordered items came from the same remote warehouse. However, following TPC-C’s default configuration, this scenario is effectively impossible. In our setup consisting of eight regions (of which seven will be remote regions), the probability of generating such a transaction is $7 \times (0.01/7)^{10} = 2.48 \times 10^{-28}$, meaning none appear among our two million samples. In contrast, real-world workloads, such as those of a global retailer, naturally exhibit such access patterns, as customers frequently place orders for items stored in remote regions. This highlights a significant gap between the synthetic behavior generated by TPC-C and that of realistic geo-distributed transaction workloads.

\para{Discussion.} In short, current evaluations only make the binary distinction between single-home and multi-home transactions~\cite{nguyen2023detock,ren2019slog,hildred2023caerus}, or do not distinguish between them at all~\cite{thomson2012calvin,shen2025mako,mu2016consolidating,harding2017evaluation,zhang2018building}. In this section, we argued that transactions should rather be split into three categories (LSH, FSH, MH). We discussed how current benchmarks should be altered to enable different scenarios (e.g., MH-heavy workloads vs. LSH-heavy workloads) for evaluating geo-distributed databases.

\begin{table}[t]
\centering
\caption{Proportions of the five transactions and three access patterns present in the default TPC-C generated data.}
\label{tab:tpcc_txn_percentages}
\vspace{-3mm}
\resizebox{\columnwidth}{!}{%
\rowcolors{2}{White}{gray!30}
\begin{tabular}{c?cccccc}
\specialrule{.1em}{.05em}{.05em}
\textbf{\begin{tabular}[c]{@{}c@{}}Transaction\\ Type\end{tabular}} & 
\textbf{\begin{tabular}[c]{@{}c@{}}\texttt{New}\\ \texttt{Order}\end{tabular}} & 
\textbf{\texttt{Payment}} & 
\textbf{\texttt{Delivery}} & 
\textbf{\begin{tabular}[c]{@{}c@{}}\texttt{Order}\\ \texttt{Status}\end{tabular}} & 
\textbf{\begin{tabular}[c]{@{}c@{}}\texttt{Stock}\\ \texttt{Level}\end{tabular}} & 
\textbf{Total} \\ \specialrule{.1em}{.05em}{.05em}
\textbf{LSH}   & 39.79\% & 43.58\% & 3.99\% & 4.00\% & 4.00\% & \textbf{95.36\%} \\ \hline
\textbf{FSH}   & 0\%     & 0.44\%  & 0\%    & 0\%    & 0\%    & \textbf{0.44\%}  \\ \hline
\textbf{MH}    & 4.21\%  & 0\%     & 0\%    & 0\%    & 0\%    & \textbf{4.21\%}  \\ \specialrule{.1em}{.05em}{.05em}
\textbf{Total} & 44.00\% & 44.02\% & 3.99\% & 4.00\% & 4.00\% & \textbf{100.00\%} \\ \specialrule{.1em}{.05em}{.05em}
\end{tabular}%
}
\end{table}
\section{Evaluation Practices in the Wild}
\label{sec:background}

Modern geo-distributed databases expose rich instrumentation, including transaction metrics, network statistics, and resource usage. Yet, existing evaluations typically focus just on latency and throughput, providing only a partial view. In this section, we analyze five representative systems, derive a taxonomy of their design choices, review their evaluation practices, and highlight limitations we resolve with \framework.

\para{System Selection.} To analyze core geo-distribution protocols under controlled conditions, we focus on five state-of-the-art open-source systems: Calvin~\cite{thomson2012calvin}, SLOG~\cite{ren2019slog}, Detock~\cite{nguyen2023detock}, Janus~\cite{mu2016consolidating}, and CockroachDB (CRDB)~\cite{taft2020cockroachdb}. Our selection is guided by the following criteria:

\begin{itemize}
    \item \textbf{Controlled comparison:} Calvin, SLOG, Detock, and Janus share a common codebase~\cite{nguyen2023detock}, enabling an apples-to-apples comparison of OLTP protocols.
    \item \textbf{Protocol diversity:} The systems span a range of approaches: deterministic ordering (Calvin), locality-aware execution (SLOG), decentralized deadlock resolution (Detock), consensus-integrated concurrency control (Janus), and optimistic concurrency control (CRDB).
    \item \textbf{Research vs.\ production:} We include CRDB to demonstrate how \framework applies to production-grade systems with strong fault-tolerance guarantees. Due to confounding factors such as proprietary query optimizers, network stacks, and security features, we focus on trends rather than direct performance comparisons.
    \item \textbf{Reproducibility:} All systems are open-source and deployable in controlled environments, enabling consistent evaluation across scenarios.
\end{itemize}

\para{Integrating Systems into \framework.} \framework is designed to be extensible. Workload generation and metric collection are decoupled from the system under test, requiring only a thin integration layer. Specifically, a system must expose executed transaction characteristics such that \framework can classify their type (SP/MP and LSH/FSH/MH). This design enables rapid integration of new protocols without modifying the core framework.

We focus on open-source systems to enable full control over deployment, instrumentation, and fault injection. Evaluating managed services such as Spanner~\cite{corbett2013spanner} or Aurora~\cite{verbitski2017amazon} introduces practical limitations (e.g., restricted observability and lack of fault injection).


\subsection{Properties of Geo-Distribution} 

From the chosen systems, we derive all key properties summarized in Table~\ref{tab:sequencer_taxonomy}. This taxonomy captures the systems' design decisions and their corresponding targeted scenarios. Replication is the most critical axis, addressing \emph{when}, \emph{how}, and \emph{where} data is replicated. \emph{(i) Mode}: Replication can be synchronous or asynchronous (some systems offer both). \emph{(ii) Scope}: Data can be replicated fully or partially in remote regions. More specifically, although only Calvin is designed for full replication, most systems have been designed for partial replication (data only replicated to some or no remote regions). Finally, \emph{(iii) Placement}: data may be geo-replicated if at least one secondary copy is placed in a remote region.

The replication strategy directly affects fault tolerance and performance. \emph{Fully synchronous} geo-replication maximizes resilience but incurs latency, as all replicas must acknowledge each transaction. Conversely, \emph{partial asynchronous} replication can yield higher throughput. Partitioning and geo-distribution are also common design choices between improving scalability and reducing latency by placing data closer to user demand.

Some systems further support \emph{dynamic home movement} (also known as remastering)~\cite{ren2019slog,nguyen2023detock,abebe2020dynamast,katsarakis2021zeus}, which adapts to changing access patterns by relocating data closer to the source of demand. This mechanism optimizes latency over time by converting FSH and MH transactions into LSH, thereby reducing the frequency of cross-regional operations. Lastly, CRDB, as the only system in \Cref{tab:sequencer_taxonomy}, implements a fault-tolerance mechanism that allows healthy parts of the cluster to continue operating even in the presence of a failure. While it trades off performance relative to other systems, CRDB offers strong resilience for users. No data will be lost, albeit at the expense of additional synchronous replication and durability mechanisms on the critical path.


\begin{table}[t]
\centering
\rowcolors{2}{White}{gray!30}
\small
\caption{System properties that refer to geo-distribution, across the five systems that we consider in this evaluation.}
\label{tab:sequencer_taxonomy}
\vspace{-3mm}
\resizebox{\columnwidth}{!}{%
\begin{tabular}{lccccccccc}
\specialrule{.1em}{.01em}{.01em}
\multirow{2}{*}{} & \multicolumn{5}{c}{\textbf{Replication}} & \multirow{2}{*}{} & \multirow{2}{*}{} & \multirow{2}{*}{} & \multirow{2}{*}{} \\ \cmidrule(lr){2-6}
\multirow{-2}{*}{\textbf{System}} & \textbf{Async} & \textbf{Sync} & \textbf{Partial} & \textbf{Full} & \textbf{Geo} & \multirow{-2}{*}{\begin{tabular}[c]{@{}c@{}}\textbf{Parti-}\\ \textbf{tioned}\end{tabular}} & \multirow{-2}{*}{\begin{tabular}[c]{@{}c@{}}\textbf{Geo-}\\ \textbf{distr.}\end{tabular}} & \multirow{-2}{*}{\begin{tabular}[c]{@{}c@{}}\textbf{Dyn. home}\\ \textbf{movement}\end{tabular}} & \multirow{-2}{*}{\begin{tabular}[c]{@{}c@{}}\textbf{Fault}\\ \textbf{Tolerance}\end{tabular}} \\ \midrule
Calvin  & \cmark & \cmark & \xmark & \cmark & \cmark & \cmark & \xmark & \xmark & \xmark \\ \hline
SLOG    & \cmark & \cmark & \cmark & \xmark & \cmark & \cmark & \cmark & \cmark & \xmark \\ \hline
Detock  & \cmark & \cmark & \cmark & \xmark & \cmark & \cmark & \cmark & \cmark & \xmark \\ \hline
Janus   & \cmark & \xmark & \cmark & \xmark & \cmark & \cmark & \cmark & \xmark & \xmark \\ \hline
CRDB    & \xmark & \cmark & \cmark & \xmark & \cmark & \cmark & \cmark & \xmark & \cmark \\ \hline
\end{tabular}
}
\end{table}

\subsection{Existing Evaluations}

We now survey the five selected database systems and analyze how their evaluations were conducted.

\para{Calvin}~\cite{thomson2012calvin} is an early deterministic OLTP system designed to support MP transactions at scale. Its core design centers on a deterministic transaction ordering layer implemented via a centralized sequencer, thereby avoiding commit-time coordination logic. Calvin does not incorporate the concept of transaction \emph{homes} and ensures strong consistency through full replication. As such, its performance remains unaffected by changes in the prevalence of FSH and MH transactions. 

\noindent \textit{Focus: } Calvin's evaluation focus is on scalability. It solely makes use of \texttt{NewOrder} transaction from TPC-C and compares exclusively against a System~R$^*$-style~\cite{mohan1986transaction} baseline.


\para{SLOG}~\cite{ren2019slog} reduces LSH transaction latency by decoupling their execution from global ordering; only MH transactions are globally ordered. This design reduces coordination overhead and improves latency in workloads dominated by LSH transactions. Experiments demonstrate that SLOG maintains throughput comparable to Calvin while significantly lowering latency for non-MH transactions.

\noindent \textit{Focus:} The original evaluation includes both synthetic YCSB workloads and \texttt{NewOrder} transactions from TPC-C. It compares against Calvin, Spanner, and a two-phase locking (2PL) baseline. Overall, the evaluation emphasizes scalability and latency improvements across varying LSH and MH ratios, and examines the impact of data locality and coordination costs.

\para{Detock}~\cite{nguyen2023detock} eliminates global sequencing by relying on a distributed, graph-based deadlock-resolution protocol that runs independently on each server. By avoiding global ordering and minimizing aborts, Detock improves the throughput and latency of MH transactions, particularly under contentious workloads. Caerus~\cite{hildred2023caerus} follows a very similar design. 

\noindent \textit{Focus:} The evaluation exercises synthetic YCSB workloads and a complete TPC-C benchmark, comparing to Calvin, SLOG, Janus, and normalized CRDB. The study emphasizes the scalability, abort-free behavior, and performance of MH transactions under asymmetric network conditions.

\begin{table}[t]
\centering
\caption{Resources and pricing of various AWS instance types~\cite{aws_pricing}.}
\label{tab:vm_instance_pricing}
\vspace{-3mm}
\resizebox{\columnwidth}{!}{%
\rowcolors{2}{White}{gray!30}
\begin{tabular}{r?cccc}
\bottomrule
\begin{tabular}[r]{@{}l@{}}\textbf{VM Instance}\\ \textbf{Type}\end{tabular}   & \textbf{vCPUs} & \begin{tabular}[c]{@{}l@{}}\textbf{Memory}\\ \textbf{(GiB)}\end{tabular}& \begin{tabular}[c]{@{}l@{}}\textbf{Network}\\ \textbf{Performance (GiB/s)}\end{tabular} & \begin{tabular}[c]{@{}l@{}}\textbf{Cost (\$/h)}\\ \textbf{in us-west-1}\end{tabular} \\ \hline
m5.2xlarge & 8 & 32 &  $\le10$ GiB & \$ 0.448 \\ \hline
r5.2xlarge & 8 & 64 &  $\le10$ GiB & \$ 0.560 \\ \hline
m5.4xlarge & 16 & 64 & $\le10$ GiB & \$ 0.896 \\ \hline
r5.4xlarge & 16 & 128 & $\le10$ GiB & \$ 1.120 \\ \hline
m6i.8xlarge & 32 & 128 &  $\le12.5$ GiB & \$ 1.792 \\ \toprule
\end{tabular}%
}
\end{table}



\para{Janus}~\cite{mu2016consolidating} unifies concurrency control and consensus to avoid cycles in the serialization graph. Leveraging their shared requirements allows most transactions to commit in a single round trip, improving performance under high contention.

\noindent \textit{Focus:} The evaluation includes YCSB-style microbenchmarks and the full TPC-C workload, and compares Janus against 2PL, optimistic concurrency control (OCC), and TAPIR. The study emphasizes scalability and performance across varying contention and workload skew, highlighting Janus’s ability to commit transactions in high-conflict settings efficiently.

\para{CRDB}~\cite{taft2020cockroachdb} was designed to offer high resilience, availability, and strong consistency for geo-distributed applications. To provide fault tolerance, CRDB synchronously replicates data across different geographical regions at the expense of extending the critical path. CRDB can thus continue to operate in the remaining regions, even in the event of a complete regional failure. Unlike deterministic protocols, CRDB aborts transactions in the event of conflicts, but it does not require knowing the read \& write set a priori.

\noindent \textit{Focus:} The authors compare CRDB against other industry products, Amazon Aurora, and Spanner using YCSB and TPC-C, concentrating primarily on scalability and fault tolerance.


\para{Discussion.} Overall, we find that existing evaluations overlook five main aspects: $i)$~monetary costs are not considered at all, $ii)$~systems are only evaluated on a single type of cloud VM, $iii)$~the evaluations are restricted to a narrow set of benchmarking workloads and scenarios, $iv)$~FSH transactions have not been considered, and $v)$~the effect of diverse network conditions is investigated insufficiently. The remainder of this paper addresses these aspects.
\section{Towards Holistic Evaluation Scenarios}
\label{sec:scenarios_metrics:eval_scenarios}

In this section, we devise three deployment scenarios: \textit{varying input throughput}, \textit{resource allocation}, and \textit{server geo-distribution}, two benchmark-specific scenarios with respect to: \textit{access patterns} and \textit{access skew}, and finally three resilience scenarios: \textit{network latency \& jitter}, \textit{packet loss}, and \textit{fault tolerance}. We also justify why these scenarios are essential for evaluating geo-distributed databases.


\subsection{Deployment-specific Scenarios}

\para{Varying Input Throughput.} As a starting point to link with other related work, we develop a scenario to measure the number of requests that a system can handle without the latency exceeding a certain threshold. Consider an online retailer with a varying number of customers throughout the day. \framework progressively increases the number of clients generating transactions at a fixed (per-client) rate to measure how well the systems can sustain the input rate and the associated latency.

\begin{table}[t]
\centering
\caption{Data and server distribution in the server geo-distribution scenario.}
\label{tab:server_skew_distributions}
\vspace{-3mm}
\resizebox{\columnwidth}{!}{%
\rowcolors{2}{White}{gray!30}
\begin{tabular}{r?cccc}
\specialrule{.1em}{.05em}{.05em}
& \multicolumn{4}{c}{\textbf{Server Geo-distribution (\%)}} \\
\cmidrule[0.06em]{2-5} 
\textbf{Configuration} & \textbf{US West} & \textbf{US East}& \textbf{EU West} & \textbf{AP NorthEast}\\ 
\specialrule{.1em}{.05em}{.05em}
uniform    & 25\%   & 25\%   & 25\%   & 25\%   \\ \hline
usw+       & 37.5\% & 25\%   & 25\%   & 12.5\% \\ \hline
usw+ eu+   & 37.5\% & 12.5\% & 37.5\% & 12.5\% \\ \hline
usw++      & 62.5\% & 12.5\% & 12.5\% & 12.5\% \\ \hline
usw++ eu++ & 50\%   & 0\%    & 50\%   & 0\%    \\ 
\specialrule{.1em}{.05em}{.05em}
\end{tabular}%
}
\end{table}

\para{Resource Allocation.} Practitioners want to optimize the needed hardware resources and costs associated with achieving the desired performance. Four key resources drive the performance of a geo-distributed OLTP database: CPU, DRAM, Disk I/O, and Network bandwidth. The growing number of hardware offerings from major cloud providers makes it challenging for practitioners to find the cheapest setup that still meets their SLO. E.g., AWS currently offers over 850 EC2 instance types~\cite{aws_pricing}. Yet, different concurrency mechanisms yield vastly different hardware requirements and the relationships with system performance are non-linear. \framework deploys systems on different cloud hardware, under identical benchmark configurations, to measure the cost-effectiveness of each system on cheaper or higher-end hardware. In this work, we showcase the systems' abilities on the subset of EC2 instance types listed in \Cref{tab:vm_instance_pricing}, providing a variety of different CPU, DRAM, and network bandwidth mixes.

\para{Server Geo-distribution.} Existing evaluations typically split data and transaction workloads equally across all regions. However, in practice, certain regions often handle more requests than others. To test the influence of different server geo-distribution scenarios, \framework assigns servers, along with their associated data and transactions, to one or two dominant regions, to observe the effect on performance. \Cref{tab:server_skew_distributions} summarizes the server distributions tested.

\subsection{Benchmark-specific Scenarios}

\para{Access Patterns.} While TPC-C and YCSB are relatively static, production workloads may require different proportions of LSH:FSH:MH transactions. For the example of an online retailer, the access pattern should resemble customers purchasing more products from different parts of the world. As the FSH and MH percentages increase, more transactions will need to be coordinated to maintain consistency, triggering more conflicts. With more frequent conflicts, more transactions will need to be aborted, reordered, or repeated, decreasing throughput and increasing latency.

\para{Access Skew.} In practice, the partitions accessed by transactions are rarely distributed uniformly. Workloads can exhibit a small set of frequently accessed records, increasing contention and potential conflicts. Moreover, these access patterns may evolve over time, e.g., in a diurnal cycle~\cite{van2024tpc}. For practitioners, it is paramount that databases remain unaffected by such behavior. Previous work proposed various skew implementations. E.g., Calvin and SLOG use a contention index, "HOT", to dictate the \emph{hot record set size}~\cite{ren2019slog,thomson2012calvin}. The evaluation of Detock instead defines "HOT" as the \emph{hot record set size reciprocal}~\cite{nguyen2023detock}. Other works instead declare a \emph{Zipfian coefficient $\theta$} between 0.0 (uniform distribution) and 1.0 (extreme skew)~\cite{harding2017evaluation,lu13aria,mu2016consolidating,zhang2018building,wang2022study,zhang2024fast}. In \framework we opt for the more widely used Zipfian coefficient option, as it does not require prior knowledge of the maximum possible hot record set size.


\subsection{Resilience Scenarios}

\para{Network Latency \& Jitter.} Most database evaluations assume ideal and static network conditions. Yet, cloud and on-premise production environments exhibit significant fluctuation, leading to increased transaction conflicts, timeouts, and performance bottlenecks. Empirical studies reveal that over two-thirds of internet paths experience $>$100ms latency, especially under load~\cite{hoiland2016measuring}. The p50 latencies between on the slowest connections on AWS reach up to 600ms~\cite{aws_latency}. Geo-distributed deployments further amplify this with heavy-tailed round-trip time distributions~\cite{sundberg2024measuring} and high jitter~\cite{underwood2018measuring}. \framework's network scenario induces configurable fixed propagation delays to simulate bandwidth overload and randomized jitter to mimic network instability and transient queueing effects. A practitioner can also use this scenario to investigate the system performance when servers are placed in regions locations closer together or wider apart. By setting the cross regional latency to 0ms, \framework can also cover a potential single-region deployment.

\para{Packet Loss.} In real-world networks, congestion or external interference makes packets not only slower, but sometimes also lost entirely. Cloud providers typically maintain packet loss under 0.1\%, yet under high load, or on some on-premise setups, users can experience significantly higher rates of 1–2\%~\cite{haq2017measuring,li2016lossradar}. Even modest packet-loss rates can degrade performance, as transactions are repeatedly blocked or aborted when critical coordination messages are lost. In the packet-loss scenario, \framework randomly drops a user-specified percentage of outgoing packets at each server.


\para{Fault Tolerance.} Cloud providers such as AWS, GCP, and Azure offer very high availability and low mean time to recovery. Nonetheless, critical business applications cannot compromise on availability even in the event of short, transient hardware failures. Hence, a resilient OLTP database must not only recover from a failure successfully without corrupting the data, but also ideally allow (most) users to continue using the database during the failure~\cite{zhang2018building,shen2025mako}. \framework evaluates a database's fault tolerance scheme by running two failure traces: \emph{i)} the failure of a single server in the cluster, and \emph{ii)} the failure of all database servers in a given region. \framework will then analyze the throughput before, during, and after the failure in the affected region, as well as in the cluster as a whole.

\section{Metrics}\label{sec:metrics}

To assess database systems in both performance and efficiency, we devise five key performance metrics in \framework: \textit{throughput}, \textit{latency}, \textit{aborts}, \textit{data transfers}, and \textit{cost}. Throughput, and latency are well-established performance metrics. Yet, in production environments, practitioners must not only meet their SLOs but also minimize hardware costs. To assess the efficiency of the different systems, we additionally include data transfer volume and cost per transaction in our evaluation.

\para{Throughput.} We measure the system-under-test's output throughput. \framework submits pregenerated transactions at a constant rate adjusted to meet the needs of our experiments. Clients work asynchronously; they do not wait for a response to the previous request before submitting a new transaction.

\para{Latency.} Latency represents the time interval between a client submitting a transaction to the server and receiving a response. In contrast to prior work, \framework distinguishes latency by transaction type (LSH, FSH, MH) for more fine-grained analysis.


\para{Aborts.} Some protocols may abort or restart a transaction, e.g., when a deadlock is found, or when it contains foreign key lookups on data changed in the meantime, like in PPS~\cite{harding2017evaluation}.

\para{Data Transfers.} For data transfers, we sum up all egress traffic from all database servers. We choose egress rather than ingress traffic because egress is specifically what major cloud providers (AWS, GCP, Azure) charge their customers for. Data transfers within the same cloud region are typically free of charge, while data transfers to another region are billed by data volume~\cite{aws_pricing,gcp_network_pricing,azure_network_pricing}. In the context of geo-distributed OLTP databases, a cost-efficient system must rely on light-weight internode communication algorithms. To measure data transfers between regions, \framework uses an adjusted version of iftop~\cite{iftop}, a lightweight packet-monitoring tool that uses packet sniffing to determine the amount of outgoing traffic.

\para{Cost per Transaction.} To calculate the monetary cost per transaction $C$, \framework models the three cost components of geo-distributed deployments: $i)$~VM rental cost $c_s$, $ii)$~inter-region data transfer cost $c_b$ (intra-region communication is typically free of charge), and 
$iii)$~durable storage cost for data backups $c_d$. These components reflect the standard billing models of major cloud providers (AWS, GCP, Azure)~\cite{aws_how_pricing_works,khan2024cloud,prakash2025efficient,khan2024costmodelling}. We assume the systems are deployed non-stop using on-demand VMs, thus the cost is independent of the transaction runtime. We use the following formula:

\begin{equation}
    C = \frac{n_s \cdot c_s + n_b \cdot c_b + n_d \cdot c_d}{t}
\end{equation}

where $n_s$ is the number of server nodes, $n_b$ the volume of inter-region data transferred, and $n_d$ the amount of durable storage used. Finally, $t$ denotes the system's output throughput.

\para{Resource Utilization.} Finally, we also report back to the user the \textit{CPU}, \textit{RAM}, \textit{network}, and \textit{disk I/O} utilization. While not directly affecting performance or cost, resource utilization metrics help practitioners identify over- or under-provisioned resources, guide appropriate hardware selection, or reduce power consumption costs in an on-premise setup.

\begin{figure}[t]
\centering
\includegraphics[width=\linewidth]{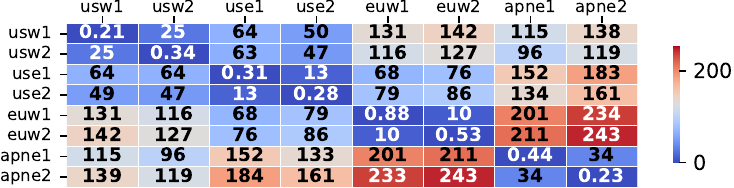}
\caption{Average round-trip times (ms) for each region pair used across 3 runs.}
\label{fig:evaluation:RTTs}
\end{figure}

\begin{figure*}[tb]
\centering
\includegraphics[width=\linewidth]{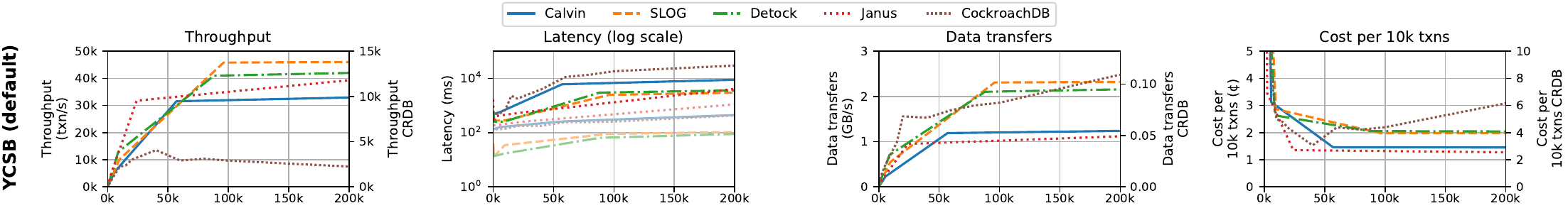}
\includegraphics[width=\linewidth]{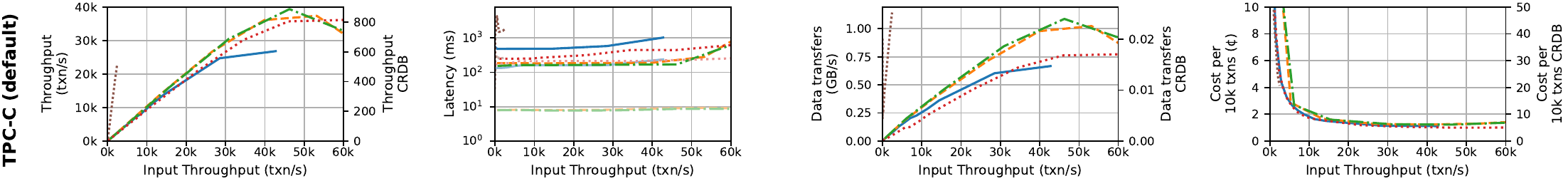}
\vspace{-6mm}
\caption{The YCSB and TPC-C throughput, latency, data transfers, and cost per transaction with increasing input throughput. Opaque lines in the latency figure represent p50 and bold ones p99.}
\label{fig:evaluation:tpcc:scalability}
\end{figure*}


\section{\framework in Action}\label{sec:eval}

We make use of \framework to evaluate geo-distributed database systems across a range of scenarios. Using the systems introduced in \Cref{sec:background}, we apply \framework to YCSB and TPC-C workloads to illustrate the types of insights it enables. Specifically, we use \framework to study:
\begin{itemize}
\item how systems behave under varying workload characteristics (e.g., throughput, access skew, and access patterns)
\item how efficiently systems utilize resources and the resulting cost implications
\item how systems respond to network instabilities and failures
\end{itemize}

Rather than direct system comparisons, our goal is analyzing how different system design choices (\Cref{sec:background}) respond to \framework's evaluation dimensions. The results are not intended as definitive system comparisons, but as illustrative case studies of how \framework exposes design trade-offs.

\subsection{Methodology}

\para{Cluster Setup.} Experiments run on \texttt{r5.4xlarge} VMs in AWS (16 vCPUs, 128 GB DRAM, $\leq$10 Gb/s bandwidth, \$1.008–1.216/h depending on the region [\texttt{us-west-1/2}, \texttt{us-east-1/2}, \texttt{eu-west-1/2}, \texttt{ap-northeast-1/2}])~\cite{aws_pricing}. Inter-region data transfers cost ¢1–9/GB~\cite{aws_pricing}. Each region hosts four database servers (as in Detock~\cite{nguyen2023detock}) and one client VM that generates transactions.

\para{Network Profile.} Measured RTTs vary significantly across regions (\Cref{fig:evaluation:RTTs}), particularly for Asia-bound traffic. AWS routing further amplifies this effect: \texttt{us-west}$\leftrightarrow$\texttt{eu-west} is 1.7$\times$ faster than \texttt{eu-west}$\leftrightarrow$\texttt{ap-northeast}, despite comparable distance~\cite{aws_networking}. Some scenarios require controlled network impairments, \framework thus utilizes NetEm~\cite{hemminger2005network}. All VMs are then co-located in a single region, using NetEm to inject delays, jitter, or packet loss based on the measured baselines.

\begin{table}[t]
  \centering
  \caption{Summary of workload family characteristics.}
  \label{tab:workload_characteristics}
  \vspace{-3mm}
  
  \resizebox{\columnwidth}{!}{%
  \rowcolors{2}{white}{gray!30}
  \begin{tabular}{l?ll}
    \toprule
    \textbf{Workload Family} & \textbf{Access Pattern} & \textbf{Skew} \\
    \midrule

    \textbf{YCSB Default} & LSH=50\%; FSH=25\%; MH=25\%  & $\theta=0.0$ \\
    \textbf{YCSB Access Patterns} & LSH=0-100\%; FSH=0-50\%; MH=0-50\%  & $\theta=0.0$\\
    \textbf{YCSB Access Skew} & LSH=50\%; FSH=25\%; MH=25\%  & $\theta=0.0-1.0$\\
    \midrule
    \textbf{TPC-C Default} & LSH=95\%; FSH=0.4\%; MH=4.2\%  & $\theta=0.0$ \\
    \textbf{TPC-C Access Patterns} & LSH=12-100\%; FSH=0-44\%; MH=0-44\%  & $\theta=0.0$\\
    \textbf{TPC-C Access Skew} & LSH=95\%; FSH=0.4\%; MH=4.2\%  & $\theta=0.0-1.0$\\
    \bottomrule
  \end{tabular}%
  } 
\end{table}

\para{Experiment Setup.} \framework supports combining multiple evaluation dimensions, but full grid search would be excessive. Instead, we adopt an ablation-style methodology to isolate the impact of each factor. \Cref{tab:workload_characteristics} summarizes the workload families used. Unless stated otherwise, we use uniform skew. For YCSB, we set FSH+MH=50\% and MP=50\%;  by default, touching 2 hot and 8 cold keys per transaction. TPC-C is configured with 1200 warehouses and a 1\% probability of a remote item order or payment.

All systems store data in memory. Calvin and Janus replicate data across all regions; SLOG and Detock store data in a primary region (their most performant system variant); and CRDB uses multi-region replication for fault tolerance. We plot the output throughput, p50 (opaque) and p99 (bold) latency, data transfers, and cost per transaction. For CRDB, we use a separate y-axis as its performance is not comparable to the other systems. For the network latency and packet loss scenarios, we normalize results relative to default conditions. For space reasons, abort rates, resource utilization, and latency ablation are only discussed in the text, with additional plots available in our GitHub repository~\cite{extended_gaia_report}.

\subsection{Varying Input Throughput}

\begin{figure*}[t]
\centering
\includegraphics[width=\linewidth]{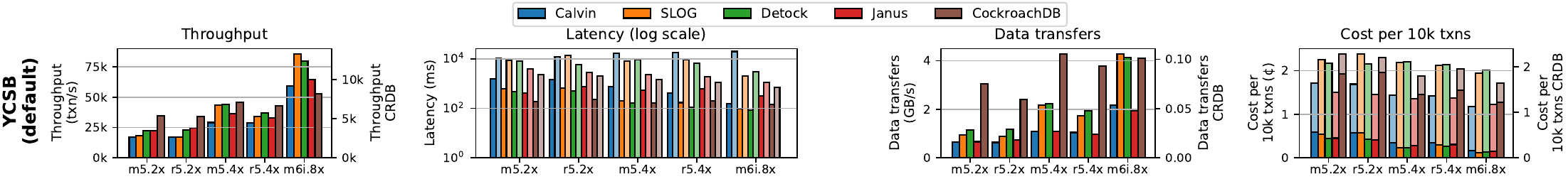}
\includegraphics[width=\linewidth]{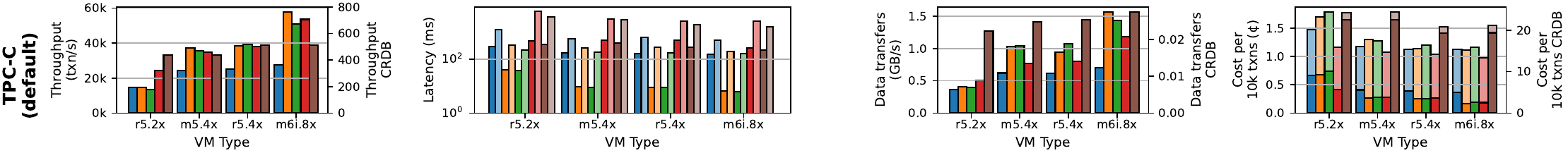}
\vspace{-7mm}
\caption{The YCSB and TPC-C throughput, p50 (bold bars) and p99 (opaque bars) latency, data transfers, and VM+storage cost (bold bars) and data transfer costs (opaque bars) per transaction across five AWS VM instance types.}
\label{fig:evaluation:tpcc:cost_heatmap}
\end{figure*}

We begin by using \framework to identify the sustainable operating throughput of each system. To this end, we vary the target input throughput from 0 to 200k (YCSB) and 60k (TPC-C) transactions/s (\Cref{fig:evaluation:tpcc:scalability}).

\para{Finding the Maximum Throughput.} As input throughput increases, all systems exhibit a characteristic saturation point beyond which latency grows rapidly. The location of this point varies per system and workload. For example, SLOG and Detock sustain higher YCSB input throughput ($\sim$100k) before their p99 latency explodes above 10s. Calvin and Janus reach their saturation points earlier in YCSB, with Janus scaling faster, but plateauing sooner. In TPC-C, memory pressure becomes the limiting factor at higher loads. Since TPC-C transactions are generally more complex than in YCSB, considerably more RAM is needed to store their in-flight data.

\para{Positioning CRDB in the Evaluation.} CRDB originates from a separate codebase, and operates in a different resilience regime than the academic prototypes. Among the evaluated systems, only CRDB is equipped with production-grade multi-region fault tolerance. In contrast, the academic research prototypes evaluated here lack support for comparable fault-tolerance guarantees. This design choice explains why CRDB achieves lower peak throughput but provides stronger resilience. CRDB's observed peak throughput (4k for YCSB and 500 for TPC-C), is consistent with the tpmC (\texttt{NewOrder} transactions per minute) results in prior reports~\cite{taft2020cockroachdb,vanbenschoten2022enabling}, accounting for cluster size, replication, row count, and geographical node placement. Closest to our setup is an experiment in~\cite{vanbenschoten2022enabling}, with TPC-C run over 10 regions, yielding $\sim$12k tpmC, corresponding to $\sim$450 txns/s. For the remaining scenarios, we thus use the peak sustainable input throughput for each system and workload. Instead of absolute-number comparisons across systems, we focus on relative performance trends.

\begin{figure*}[t]
\centering
\includegraphics[width=\linewidth]{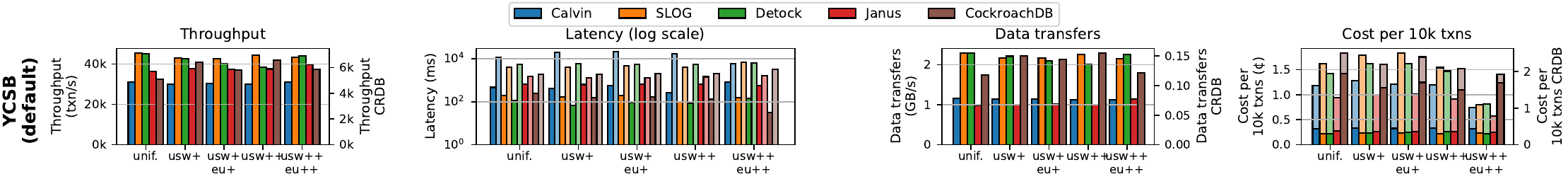}
\includegraphics[width=\linewidth]{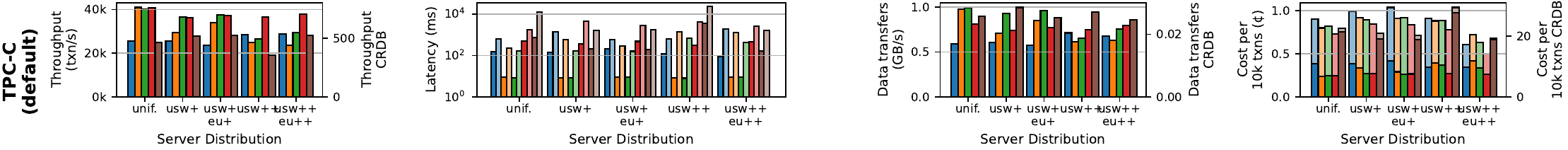}
\vspace{-7mm}
\caption{The YCSB and TPC-C throughput, p50 (bold bars) and p99 (opaque bars) latency, data transfers, and VM+storage cost (bold bars) and data transfer costs (opaque bars) per transaction across five different server skew configurations.}
\label{fig:evaluation:tpcc:server_skew}
\end{figure*}

\para{Data Transfers.} Throughput strongly correlates with data transfers. This is because increased transactional throughput leads to more communication demand for coordination, conflict resolution, and replication. Although the YSCB and TPC-C transactions themselves are relatively small, \framework measures the entire system traffic, which surprisingly is on the order of GB/s. \Cref{fig:evaluation:tpcc:scalability} also shows that Calvin's and Janus' concurrency protocols generally require less communication per transaction than SLOG and Detock.

\para{Cost.} The monetary costs per transaction are directly related to the throughput and data transfers. VM rental and durable storage have a fixed cost of around 1¢. The higher the throughput, the lower the fixed cost per transaction. When the systems reach their stabilized maximum throughput, the total cost per transaction ranges from 1.5¢ per 10k transactions (Janus and Calvin) to 2.5¢ (SLOG, Detock, and CRDB) in YCSB. Since data transfers make up the majority of costs, and with SLOG and Detock requiring almost double the data transfers, their cost per transaction is higher. Interestingly, in TPC-C, Detock and SLOG require only 25\% more data transfers; the total cost for the first four systems converges to around 1.6¢ per 10k  transactions. CRDB, with a much lower throughput, costs around 20¢ per 10k transactions. Note, however, that CRDB could likely achieve similar performance on weaker machines with a lower VM rental price.

\mybox{\textbf{Takeaways:} SLOG and Detock scale the best in terms of throughput and latency. Janus' performance is almost on par, but remains the least expensive, making a great tradeoff between performance and running costs. CRDB's performance is comparatively lower, due to its hardened fault-tolerance and its support for interactive transactions.}

\begin{figure*}[tb]
\centering
\includegraphics[width=\linewidth]{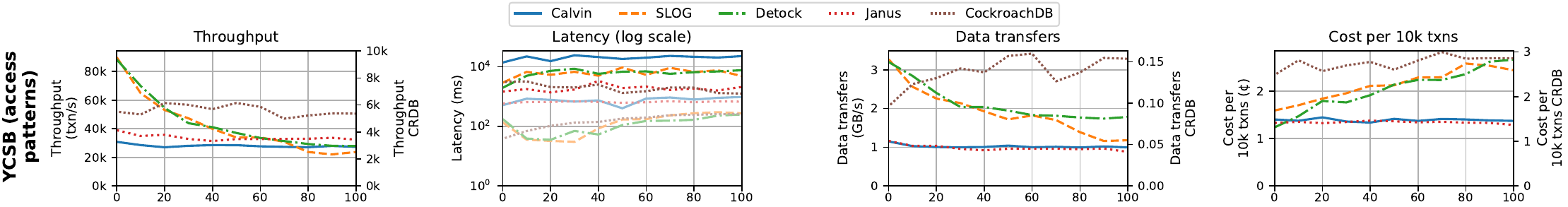}
\includegraphics[width=\linewidth]{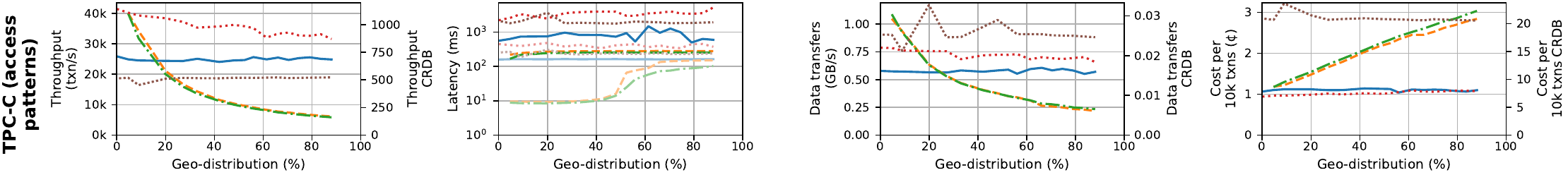}
\caption{The YCSB and TPC-C throughput, latency, data transfers, and cost per transaction with varying geo-distribution percentages. Opaque lines in the latency figure represent p50 and bold ones p99.}
\label{fig:evaluation:tpcc:normal_conditions}
\end{figure*}

\subsection{Resource Allocation} 

Next, we use \framework to examine how system behavior changes across different hardware configurations (\Cref{tab:vm_instance_pricing}) under the default YCSB and TPC-C workloads. We report monetary cost by separating fixed VM \& storage cost (bottom component) from variable data transfer cost (upper component).

\para{VM Price, Performance, and Cost.} In \Cref{fig:evaluation:tpcc:cost_heatmap}, we order machine types by available CPU and RAM from weakest to strongest. We omit the \texttt{m5.2x} machine type for TPC-C, as it has insufficient memory and crashes. Across systems, increasing hardware capacity generally improves throughput, although not strictly monotonically (e.g., \texttt{m5.4x} occasionally outperforms the more powerful \texttt{r5.4x}). 

A key observation enabled by \framework is that more powerful machines reduce the cost per transaction despite higher hourly prices. The higher attainable throughput amortizes fixed infrastructure costs and shortens the critical path of transactions. This effect is consistent across systems, although varying in magnitude. The cost breakdown further reveals different bottlenecks: for Calvin, SLOG, Detock, and Janus, data transfer dominates total cost, while CRDB spends a larger fraction on compute resources. Correspondingly, more powerful machines generally reduce latency through improved parallelism. The impact is most pronounced for SLOG and Detock in YCSB, while in TPC-C, Calvin and Janus benefit the most.

For the remaining experiments, we use \texttt{r5.4x} instances, consistent with prior work~\cite{nguyen2023detock},  allowing for direct comparisons.

\mybox{\textbf{Takeaways:} In short, increasing hardware capacity can reduce cost per transaction by improving throughput and amortizing fixed costs. \framework's resource allocation scenario highlights that, for most systems, inter-region data transfer dominates overall cost, rather than compute and storage.}

\subsection{Server Geo-distribution}

To examine the impact of the geographical server distribution, \framework evaluates the systems' performance in five different server distribution settings in \Cref{fig:evaluation:tpcc:server_skew}. Across systems, we observe a negligible impact on throughput, latency, and data transfers. This behavior is expected since the access patterns of the workloads remain unchanged. Despite clustering data in hotspots, the LSH:FSH:MH ratio remains the same.

However, there is a clear impact on cost. In highly skewed geo-distribution (\texttt{usw++eu++}), more data transfers become intra-regional, not billed by cloud providers. Thanks to the reduced inter-region communication, the cost per transaction for \texttt{usw++eu++} is up to 2x cheaper than in \texttt{uniform}.

\begin{figure*}[tb]
\centering
\includegraphics[width=\linewidth]{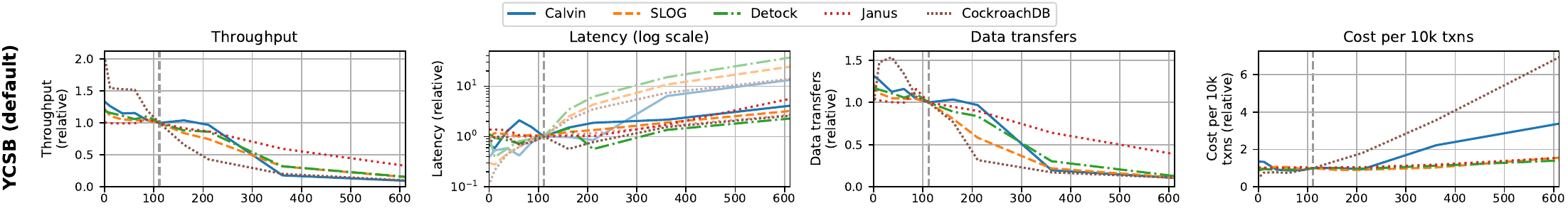}
\includegraphics[width=\linewidth]{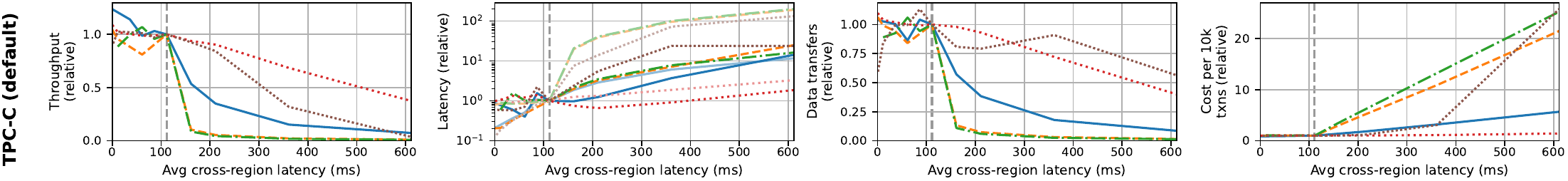}
\caption{The relative YCSB and TPC-C throughput, latency, data transfers, and cost per transaction under extra network latency and jitter. Opaque lines in the latency figure represent p50 and bold ones p99. The dashed vertical line represents the average cross-region latency in the default geo-distribution setup.}
\label{fig:evaluation:tpcc:slow_network}
\end{figure*}

\subsection{Access Patterns}

While YCSB natively supports varying the transaction access patterns, TPC-C defines a fixed LSH:FSH:MH transaction ratio. To provide such a test on TPC-C, \framework varies the probability of targeting a remote warehouse in a \texttt{NewOrder} or \texttt{Payment} transaction. In this manner, our \emph{TPC-C Access Patterns} workload controls the geo-distribution percentage from 0\% to 88\%, making the workload increasingly harder for the systems. For YCSB, a geo-distribution of 100\% means having 0\% LSH, 50\% FSH, and 50\% MH transactions. In the case of TPC-C, the maximum geo-distribution percentage of 88\% is decomposed into 44\% FSH and 44\% MH transactions; the remaining 12\% of transactions can only be LSH by design.

\Cref{fig:evaluation:tpcc:normal_conditions} illustrates the diverse effects of increasing geo-distribution depending on system design. SLOG and Detock experience a steady decline in throughput as the cross-region share increases, accompanied by a 40\% drop in CPU utilization. In contrast, Calvin, Janus, and CRDB remain unaffected, because they treat LSH, FSH, and MH transactions equally.

\para{Latency.} These differences are further reflected in latency. Calvin, Janus, and CRDB maintain relatively stable latency across geo-distribution levels, albeit at consistently higher levels. In contrast, SLOG and Detock show a significant increase in p50 latency as the workload becomes more geo-distributed, particularly for TPC-C. Their initial advantage stems from optimized handling of LSH transactions, but this benefit diminishes at 100\% geo-distribution, once most transactions require cross-region coordination. As a consequence, Janus and Calvin also prove to be more cost-efficient as the workloads become more geo-distributed.

\para{Latency Ablations.} For a deeper analysis we isolate the latency of read vs. write and LSH vs. FSH vs. MH transactions. Again, Calvin and Janus do not differentiate between these transaction types, and their latency always remains high. However, \framework reveals that SLOG and Detock are an order of magnitude more efficient for read and LSH transactions. Interestingly, the latency of FSH transactions grows from 20ms up to 250ms as geo-distribution increases and conflicts with MH transactions become more frequent. CRDB also offers faster response times for reads than writes. However, LSH transactions remain slow because they still need to be replicated synchronously to a remote replica.

\mybox{\textbf{Takeaways:} SLOG and Detock optimize for LSH transactions, yet at higher geo-distribution percentages, their performance deteriorates and costs grow. Similarly, a highly fault-tolerant system like CRDB can perform well for local read transactions, but persisting writes to foreign-region replicas takes a toll on its performance. This entails that hybrid solutions are required. Janus strikes a reasonable performance-cost balance.}


\begin{figure*}[tb]
\centering
\includegraphics[width=\linewidth]{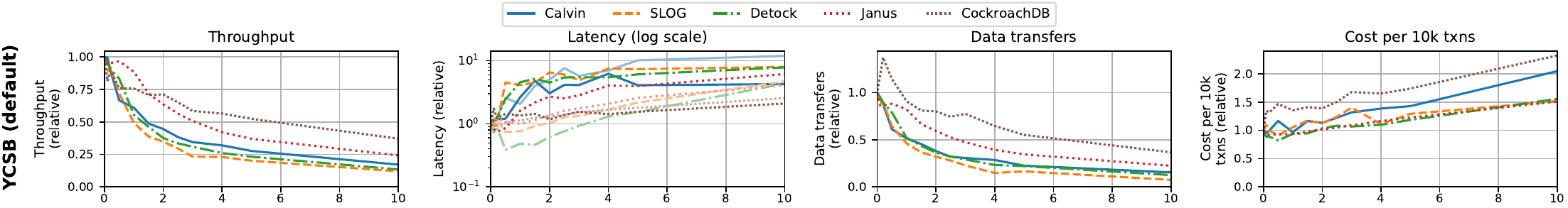}
\includegraphics[width=\linewidth]{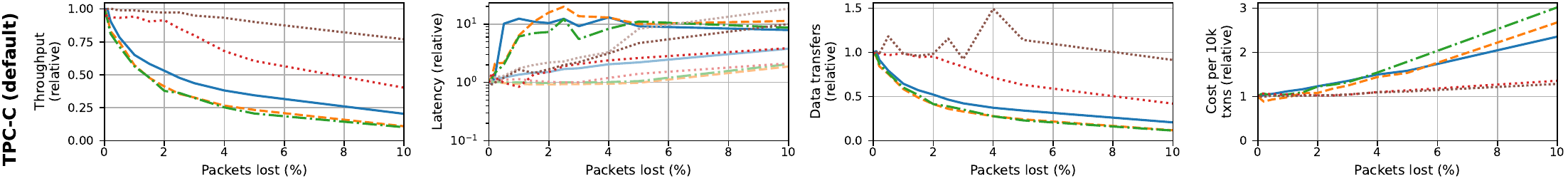}
\caption{The relative YCSB and TPC-C throughput, latency, data transfers, and cost per transaction under extra packet loss. Opaque lines in the latency figure represent p50 and bold ones p99.}
\label{fig:evaluation:tpcc:packet_loss}
\end{figure*}

\subsection{Access Skew}

In the YCSB workload, \framework defines hot and cold keys in the table. The workload's built-in $\theta$ parameter makes it suitable for evaluating access skew scenarios. A skew factor close to~0 results in a uniform distribution of all keys, whereas values close to~1 concentrate transactions on a handful of hot keys. Our \emph{TPC-C access skew}, models skew by limiting the pool of \texttt{items} and \texttt{customers} involved in the \texttt{NewOrder} and \texttt{Payment} transactions. The smaller the pool of \texttt{items} and \texttt{customers}, the more contention in TPC-C. Our results are consistent with the latest literature~\cite{nguyen2023detock}, i.e., access skew does not affect the throughput and latency of deterministic databases considerably for our selected MP configuration. The only system that exhibits a notable impact is CRDB, where extremely high contention rates trigger frequent aborts, leaving very little useful work. For brevity, we omit the plots.

\subsection{Network Latency \& Jitter}

To test the effect of diverse network conditions, \framework varies the average cross-region latency from 0ms (single-region deployment) to over 600ms (worst-case round trip time observed in AWS~\cite{aws_latency}). We also induce jitter of 10\% of that average latency. This allows \framework to identify breaking points, i.e., at what latencies does system behavior shift sharply. \Cref{fig:evaluation:tpcc:slow_network} depicts the relative performance compared to an average latency 111ms, our default geo-distributed setting. This scenario is workload-agnostic and can be applied to YCSB, TPC-C, or custom benchmarks.

Increasing latency exposes clear differences in how systems handle coordination. Janus exhibits almost no performance effect moving from 0ms to 111ms average cross region delay, and afterwards only degrades linearly. Even during the highest slowdowns, Janus sustains a meaningful amount of throughput. In comparison, SLOG and Detock degrade much more rapidly, reaching their breaking points at just $\sim$150ms average delay in TPC-C. The systems lose 90\% of their throughput and increase their latency tenfold. Similar trends are observed in YCSB with all systems except Janus quickly dropping in throughput. We attribute the robustness of Janus to high latency to the fact that it can commit non-conflicting transactions in a single round-trip. As a consequence, the high latency affects the commit time of non-conflicting transactions less than in other systems. \framework also uncovers which systems thrive in single region deployments. In particular, CRDB is able to double its YCSB throughput compared to our default setup.

\para{Cost Implications.} At high latencies, cost per transaction explodes for most systems. This is because very little useful work is still being done and the fixed infrastructure costs are amortized over fewer completed transactions. Systems that maintain some meaningful throughput under high latency conditions are therefore able to sustain lower cost per transaction.

\mybox{\textbf{Takeaways:} \framework reveals that none of the systems copes with network slowdowns effectively. Out of the five systems, Janus is the most robust and can keep the cost per transaction low, while SLOG and Detock struggle the most.}

\subsection{Packet Loss} 

We further evaluate robustness under unreliable networks by introducing 0--10\% packet loss across all connections (\Cref{fig:evaluation:tpcc:packet_loss}). Evidently, even modest packet loss (e.g., 2\%) leads to noticeable 35--60\% throughput decline and latency deterioration of Calvin, SLOG, and Detock. Only CRDB and, to some extent, Janus show more resilience to packet loss. As in the latency experiments, data transfer volume decreases with throughput, while cost per transaction increases due to reduced system utilization.

\mybox{\textbf{Takeaways:} In summary, even modest packet loss can trigger major slowdowns in all systems, although CRDB proves to be the most resilient. With extreme packet loss, latency, and costs explode, even with reduced data transferrs.}

\subsection{Fault Tolerance} 

\begin{figure}[t]
\centering
\includegraphics[width=\linewidth]{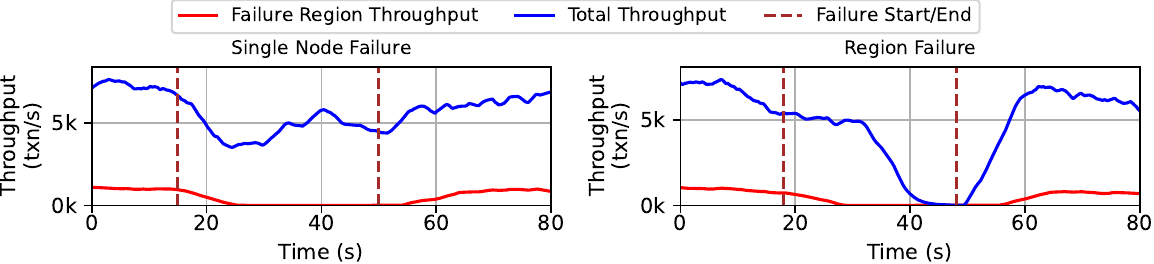}
\caption{Trace of CRDB's performance under a single node and full region failure scenario. The vertical lines indicate the start of a failure, and recovery.}
\label{fig:evaluation:ycsb:fault_tolerance}
\end{figure}

Finally, as an ultimate test of database resilience, we plot two failure throughput traces (\Cref{fig:evaluation:ycsb:fault_tolerance}). We begin by warming up the cluster for $\sim$15s to reach a steady throughput of transactions and then kill a single node or an entire region. After another $\sim$30s, the node is restarted again. Among the evaluated systems, only CRDB was actually able to sustain such failures and recover correctly. The remaining systems require manual intervention and a restart to resume operation. Furthermore, thanks to the chosen replication policy (1 replica in a foreign region), CRDB can also maintain some level of operation for clients from unaffected regions during the failure.

\para{Single Node Failure.} When a single node (out of 32) fails, CRDB's overall throughput drops by $\sim$40\%. Clients in the affected region are temporarily unable to submit transactions, although for a few seconds, it may continue to receive responses to previous requests from other nodes. The remaining nodes must adjust to the missing 3rd replica for certain data items. After the node recovery recovery, the system requires another 10s to restore normal throughput and catch up with pending requests during downtime.

\para{Region Failure.} In a region failure, some data items only have one replica online, which is insufficient for a majority. Initially, $\sim$15s after the failure, clients in other regions can continue to execute transactions touching data in unaffected regions. Eventually, however, all clients  block, waiting on unavailable partitions and bring the entire cluster to a standstill. Nonetheless, CRDB can still safely recover once the region comes back online. Its meticulous fault tolerance protocol prevents data loss, and blocked transactions can restart once the entire cluster is back in operation.

\mybox{\textbf{Takeaways:} A sound fault tolerance mechanism is paramount for resilience. While impeding performance under normal conditions, CRDB's fault-tolerance protocol ensures continued operation during a single-node failure and prevents data loss in a complete regional outage.}




\section{Lessons Learned}\label{sec:discussion}

Our research yields several interesting insights into system design, benchmarking methodology, and open challenges in geo-distributed OLTP research.

\para{Robustness of Transaction Protocols.} Protocol design plays a critical role when systems operate under network instability. Protocols with fewer coordination rounds (e.g., Janus in our testbed) prove more robust. A broader concern is that fault-tolerance mechanisms introduce non-trivial overhead to \textit{end-to-end} system performance. When evaluating concurrency control and transaction commit protocols in isolation and without fault-tolerance mechanisms in place, their latency and throughput characteristics may appear favorable. However, these advantages can degrade significantly once fault tolerance is introduced. Our experience with CRDB reinforces this point: the throughput of a production grade, fault-tolerant system is substantially lower than that of academic prototypes. We advocate for evaluating new protocols in systems where fault tolerance is a first-class concern, not an afterthought.

\para{Transfer Costs: The New Bottleneck.} Designing database systems around cost requires a fundamentally different perspective than optimizing for latency. High WAN latency often leaves RAM and disk resources underutilized, while data transfers dominate the operational bill. This tension calls for protocols that strike a balance between performance and cost efficiency, instead of optimizing for one dimension in isolation. 

This observation complements recent findings in OLTP research, which identify communication overhead as the primary performance bottleneck~\cite{zhou2025oltp}. While prior work emphasizes latency, our results show that in geo-distributed settings, communication also dominates monetary cost.


\para{Multi-Parameter Optimization.} \framework focuses on throughput, latency, data transfers, and cost per transaction. Throughput and latency often exhibit non-linear and system-specific behavior. If multiple systems satisfy an SLO, non-trivial tradeoffs across performance, cost, and availability remain. This makes multi-parameter optimization particularly challenging. Frameworks like \framework can empower architects to make well-informed decisions, uncovering a potential Pareto frontier. 


\para{Towards New Benchmarks.} Production geo-distributed workloads may diverge sharply from TPC-C and YCSB~\cite{van2024tpc,van2023cloud,zhang2023cdsben}. They feature bursty loads, long-tailed access distributions, diurnal cycles, and transaction mixes unlike today’s benchmarks (e.g., FSH transactions are nearly absent in TPC-C). Tunable workload generators~\cite{li2019lauca} move in the right direction. Yet, an in-depth study of transactions included and omitted from benchmarks such as YCSB, TPC-C, SmallBank~\cite{alomari2008cost}, MovR~\cite{cockroach2010movr}, or PPS~\cite{serafini2016clay} is needed. Based on such an analysis, future work could produce a configurable, TPC-C-like benchmark with recently observed traits by major cloud providers, similar to recent OLAP benchmarks \cite{redshiftfleet, snowflakefleet}.
\section{Related Work}\label{sec:related_work}

\Cref{sec:background} introduced the studied systems. Here, we highlight related evaluations and surveys, grouped by theme.

\para{Concurrency Control Evaluations.} The work most closely aligned with ours is~\cite{harding2017evaluation}, which found that existing protocols scale poorly in distributed, but single-region environments. Our work extends this line of inquiry to an eight-datacenter deployment spanning three continents, where WAN latencies dominate. Yu et al. evaluated seven concurrency control schemes on many-core architectures~\cite{yu2014staring}, advocating that database designs must be fundamentally rethought for novel hardware. At a finer granularity,~\cite{harizopoulos2018oltp} analyzed the instruction-level sources of OLTP latency, and~\cite{vanbenschoten2022enabling} performed a deep dive into CRDB's performance for geo-distributed databases.

\para{System Configuration and Methodology.} Leis et al. and Chang et al. investigated cost-optimal instance selection~\cite{leis2021towards,chang2025eva}; however, geo-distributed deployments additionally require accounting for high network delays. Wang et al. examined the reproducibility of recent database evaluations~\cite{wang2022study}, focusing primarily on throughput and a narrow set of scenarios. Finally, Nguyen et al. surveyed the OLTP system assumptions about the applications using them~\cite{nguyen2025database}.

\para{Benchmarking Beyond TPC-C.} To align evaluations with production settings, Poess proposed TPC-DI~\cite{poess2014tpc}, linking transactional with data integration workflows, e.g., bulk loading and transformation tasks. While valuable for mixed workloads, the adoption of TPC-DI remains limited in the community. Similarly, TPC-E~\cite{tpc2010tpc}, designed as a more sophisticated successor to TPC-C, has yet to attract mainstream adoption.
\section{Conclusion}\label{sec:conclusion}

Our study demonstrates that performance-only evaluations are insufficient for geo-distributed database systems. Network instability, inter-region data transfer, and fault-tolerance mechanisms fundamentally shape both system performance and economic feasibility in real-world deployments. Therefore, we argue that a realistic geo-distributed database evaluation must explicitly incorporate: $i)$~Local Single-Home / Foreign Single-Home / Multi Home workload composition, $ii)$~inter-region data transfer volume and cost, $iii)$~network delay, jitter, and packet loss, and $iv)$~fault tolerance under node and regional failures. Only by evaluating systems across all these dimensions can we obtain results that reflect practical cloud deployments. We believe \framework will encourage the community to embrace cost-, network-, and failure-aware benchmarking as a standard practice in geo-distributed database research.

\bibliographystyle{IEEEtran} 
\bibliography{references}    


\end{document}